\documentclass[useAMS,usenatbib,usegraphicx]{mn2e}

\usepackage{color}
\title[GRB 070518: an optically dim burst]
   {GRB 070518: A Gamma-ray Burst with Optically Dim Luminosity}

\author[Xin et al.]
{L. P. Xin$^{1,2}$\thanks{email: xlp@bao.ac.cn},
                 W. K. Zheng $^{1,2}$, J. Wang$^{1}$, J. S. Deng$^{1}$,
\newauthor
Y. Urata$^{3,4,5}$, Y. L. Qiu$^{1}$, K. Y. Huang$^{3}$,  J. Y. Hu$^{1}$, and J. Y. Wei$^{1}$\\
\\
$^1$ National Astronomical Observatories, Chinese Academy of Sciences, Beijing 100012, China.\\
$^2$ Graduate School of Chinese Academy of Sciences, Beijing 100049, China.\\
$^3$ Academia Sinica Institute of Astronomy and Astrophysics, Taipei
106, Taiwan.\\
$^4$ Department of Physics, Saitama University, Shimo-Okubo, Saitama, 338-8570, Japan.\\
$^5$ Institute of Astronomy, National Central University, Chung-Li 32054, Taiwan, Republic of China.
}

\begin{document}
\label{firstpage}

\date{Accepted 2009 Septermber 28. Received 2009 Septermber 27; in original form 2009 May 20}
\pagerange{\pageref{firstpage}--\pageref{lastpage}}
\pubyear{2009}
\maketitle

\begin{abstract}
We present our optical observations of {\em Swift} GRB 070518 afterglow obtained at the
0.8-m Tsinghua University-National Astronomical Observatory of China telescope (TNT) at Xinglong Observatory. Our follow-up
observations were performed from 512 sec after the burst trigger.
With the upper limit of redshift$\sim$0.7, GRB 070518 is found to be an
optically dim burst. The spectra indices $\beta_{ox}$ of optical to
X-ray are slightly larger than 0.5, which implies the burst might be
a dark burst. The extinction $A_{V}$ of the host galaxy is 3.2 mag
inferred from the X-ray hydrogen column density with Galactic
extinction law, and 0.3 mag with SMC extinction law. Also, it is
similar to three other low-redshift optically dim bursts,which
belong to XRR or XRF, and mid-term duration($T_{90}<10$, except
for GRB 070419A, $T_{90}$=116s). Moreover, its $R$ band afterglow
flux is well fitted by a single power-law with an index of 0.87. The
optical afterglow and the X-ray afterglow in the normal segment
might have the same mechanism, as they are consistent with the
prediction of the classical external shock model. Besides, GRB
070518 agrees with Amati relation under reasonable assumptions.
The Ghirlanda relation is also tested with the burst.
\end{abstract}

\begin{keywords}
gamma rays:bursts--gamma rays: observations--individual: GRB 070518.
\end{keywords}

\section{Introduction}
Since the first optical counterpart of gamma-ray burst (GRB) was
identified on 1997 February 28,  more and more GRBs have been optically
detected, especially after {\em Swift} was launched successfully,
which allows early follow-up observations with ground-based optical
telescopes. However, about 40 per cent of {\em Swift} GRBs still failed to
be identified with optical or infrared counterparts (Burrows et al. 2008).
For the nature of the dark burst, several models have been proposed:
extinction from dusty opaque star-forming regions ( e.g. Groot et
al. 1998; Reichart \& Price 2002); high redshift, which diminishes the GRB
optical luminosity (Lamb \& Reichart 2000), as the ly$\alpha$
forest is shifted into the observational energy range; intrinsically
dim bursts (Fynbo et al. 2001; Rol et al. 2005 ); the observational
effect, As a result of the late response to the bursts. Jakobsson et
al. (2004) have proposed an operational definition of dark burst (i.e.
optical-to-X-ray spectral index $\beta_{\mathrm{ox}}<0.5$), according
to the fireball model.

However, GRBs with an optical afterglow have been studied
widely and in depth(e.g. GRB 060218, GRB 080319B). This has resulted in 
an increasing
number of GRBs with known redshift, which has allowed statistical studies to be carried out.
The optical luminosity distribution of GRBs varies
widely up to several magnitudes (Kann et al. 2006; 2008). After
being transformed into a common redshift (e.g., $z=1$), the optical
luminosity shows a bi-model phenomenon (Nardini et al. 2006; Liang
et al. 2006; Kann et al. 2006; 2008): optically luminous and optically
dim. The bi-model distribution is also found to exist in other
energy bands (Gendre et al. 2008a,b).

Liang \& Zhang (2006) found that
the apparent bimodality cannot be interpreted as a
manifestation of the extinction effect, and they suggested that
there might be two types of progenitors or two types of
explosion mechanisms in operation.
Considering the dark burst is usually a type of GRBs with a faint optical afterglow,
optically dim bursts might be related to dark bursts (Nardini
et al. 2006). However, the number of well-studied optically dim
bursts is very limited.
Thus, increasing the number of known optically dim bursts
may help us to understand the origin of the bi-model
distribution, the nature of optically dim bursts
and dark bursts.

One such interesting case is GRB 070518.
It was detected by
Gamma-ray Burst Alert Telescope (BAT) onboard
{\em Swift} at 14:26:21 (UT) on 2007 May 18.
The X-ray telescope (XRT) and ultraviolent/optical telescope
(UVOT) on board {\em Swift} observed the counterpart
beginning at 70 and 100 sec after the trigger, respectively.
T$_{90}$ (15-350 KeV) is 5.5$\pm$0.2 s ,T$_{50}$ is 2.9s
(Sakamoto et al. 2008a).  The values of duration mean
the burst belongs to the short tail of the long group
(Kouveliotou et al. 1993). Thus, it might be a classical
long-duration GRB or an intermediate-long GRB (Horvath et al. 2008).
The ratio of fluence between the 25-50 KeV and 50-100 KeV band
(Sakamoto et al. 2008a) is about 1.06, which makes GRB 070518
belong to an X-ray rich burst (XRR), according to the criterion
\footnote{S(25-50KeV)/S(50-100KeV)$=<$0.72
C-GRB; \\ 0.72$<$S(25-50KeV)/S(50-100KeV)$=<$1.31 XRR;\\
S(25-50KeV)/S(50-100KeV)$>$1.32 XRF.} given by Sakamoto et al.
(2008b). The first given magnitude and coordinates of the
counterpart are about 18 mag in $white$ band reported by UVOT and
RA(J2000) = 16h 56m 47.7s, Dec(J2000) = 55d 17m 42.3s (radius,
90 pen cent containment), respectively (Guidorzi et al. 2007). The afterglow was
also observed by other ground-based telescopes later.
Besides, the redshift of GRB 070518 was reported to be lower than
0.7 (Cucchiara et al. 2007) based on the photometry of the {\em
Swift} UVOT. No other report about redshift of GRB 070518 has been
presented in the literature up to now. In our analysis, the redshift
of 0.7 are used to investigate the properties of GRB 070518.

In this paper, we report on optical photometric follow-up
observations of GRB 070518 at Xinglong Observatory
of National Astronomical Observatories, Chinese
Academy of Sciences.
Afterglow observations and results are described in $\S$2.
An anaysis and discussion are presented in $\S$3, 
and a summary and conclusion are
given in $\S$4.  In our calculations,
a flat universe is assumed with matter density $\Omega_M = 0.3$,
cosmological constant $\Omega_\Lambda=0.7$, and Hubble constant
$H_0=70$ km s$^{-1}$ Mpc$^{-1}$. The formula
$f\propto$$t^{-\alpha}\nu^{-\beta}$ is used for the afterglow
decay analysis.

\section{ TNT Observations and Data Reductions}

We carried out a follow-up observation programme of {\em Swift} GRBs with
the 0.8-m Tsinghua University-National Astronomical observatory of China optical telescope 
(TNT) at Xinglong Observatory,
under the framework of the East-Asia Follow-up Observation Network (EAFON; Urata et al. 2003;Urata et al. 2005).
The TNT telescope is equipped with a PI 1300$\times$1340 CCD and filters
in the standard Johnson-Bessel system.
The field of view of the TNT is 11.4$\times$11.4 arcmin,
yielding 0.5 arcsec of the pixel scale.
A custom-designed automation system
has been developed for the GRB follow-up observations (Zheng et al. 2008).

The follow-up observations of GRB 070518 were performed with the TNT from
512s after the initial {\em Swift} trigger, and lasted for about
5.5 hours until the dawn time.
A sequence of $white$, $R$, $V$, and $I$ bands images
was obtained.
After preliminary analysis,
we first confirmed the counterpart of GRB 070518 (Xin et al. 2007)
after the report of {\em Swift} UVOT.

Data reduction was carried out following standard
routine in IRAF \footnote{IRAF is distributed by NOAO, which is
operated by AURA, Inc., under cooperative agreement with NSF.}
package, including bias and flat-field corrections. Dark correction
was not performed, as the temperature of our CCD was cooled
down to $-110\,^{\circ}\mathrm{C}$. Point spread function (PSF) photometry
was applied via DAOPHOT task in IRAF package to obtain the instrumental
magnitudes. During the reduction, some frames were combined in
order to increase  signal-to-noise (S/N) ratio.
Flux calibration was performed by re-observing the field of
GRB 070518, together with Landolt photometric standard
stars with the TNT telescope on 2007 November 14 and  December 12.
Considering that the $white$ band is not a standard band,
we treated the $white$-band filter as a very broad filter.
When we calibrated the $white$ band data to $R$ band magnitude,
no significant difference was found in the colour terms,
and the response functions of
the two filters were very similar with about 0.07 mag
uncertainty.
Since the redshift of GRB 070518 is no larger than 1,
the $white$ band data was calibrated to $R$ band magnitude,
for simplicity.
The transformation uncertainty of 0.07 mag is added
to the final results.

In addition, other optical photometries for GRB 070518 are
collected from the GRB Coordinate Network (GCN), in
order to complete the optical light curves. All of these optical data are
presented in Table 1.


\begin{table}
\caption[]{A log for the optical afterglow photometry of GRB 070518. The
reference time T0 is 14:26:21 (UT) on 2007 May 18. The TNT $white$
band ($W$) magnitudes were calibrated against the $R$ band with the
uncertainty of 0.07 magnitude.
No data were corrected for the Galactic extinction of
$E_{B-V}=0.017$. The flux contamination and extinction of host
galaxy were also not corrected.}
  \label{Tab:publ-works}
  \begin{center}\begin{tabular}{ccccccc}
  \hline\noalign{\smallskip}
T-T0 &  Exposure     & Band & Mag & Telescope\\
(mid day) & (Sec) &      &     &   or Reference\\
  \hline\noalign{\smallskip}
  \hline\noalign{\smallskip}
0.00636 & 2$\times$20 &$W$ & 19.03$\pm$0.10 & TNT \\
0.00699 & 2$\times$20 &$W$ & 19.23$\pm$0.10 & TNT \\
0.00804 & 3$\times$20 &$W$ & 19.37$\pm$0.12 & TNT \\
0.00883 & 3$\times$20 &$W$ & 19.27$\pm$0.08 & TNT \\
0.00961 & 3$\times$20 &$W$ & 19.68$\pm$0.14 & TNT \\
0.01053 & 4$\times$20 &$W$ & 19.56$\pm$0.12 & TNT \\
0.01157 & 4$\times$20 &$W$ & 19.74$\pm$0.13 & TNT \\
0.01275 & 5$\times$20 &$W$ & 20.08$\pm$0.15 & TNT \\
0.01406 & 5$\times$20 &$W$ & 20.06$\pm$0.18 & TNT \\
0.02446 & 1$\times$300 &$R$ & 20.20$\pm$0.10 & TNT \\
0.02814 & 1$\times$300 &$R$ & 20.76$\pm$0.16 & TNT \\
0.03723 & 3$\times$300 &$R$ & 20.48$\pm$0.10 & TNT \\
0.01237   & - &  $R$     &    19.5      &         GCN6443  \\
0.23337   & - &  $R$     &    22        &         GCN6483  \\
0.35948   & - &  $R$     &    22.2      &         GCN6418  \\
0.37753   & - &  $Rc$    &    22.3$\pm$0.2    &         GCN6458  \\
0.71837   & - &  $R$    &    23.03$\pm$0.05  &         GCN6462  \\
1.45833   & - &  $R$     &    23.3      &         GCN6426  \\
1.70837   & - &  $R$    &    23.36$\pm$0.05  &         GCN6462  \\
3.60837   & - &  $R$    &    23.56$\pm$0.05  &         GCN6462  \\
0.02065 & 1$\times$300 &$V$ & 20.74$\pm$0.15 & TNT \\
0.06383 & 10$\times$300 &$I$ & 20.47$\pm$0.17 & TNT \\
0.10197 & 10$\times$300 &$I$ & 20.92$\pm$0.13 & TNT \\
0.15271 & 14$\times$300 &$I$ & 21.37$\pm$0.14 & TNT \\
0.20660 & 15$\times$300 &$I$ & 21.23$\pm$0.24 & TNT \\
  \noalign{\smallskip}\hline
  \end{tabular}\end{center}
\end{table}

\section{Analysis and Discussion}

\subsection{Light Curves and Chromatic Decay}

The Multiband light curves of GRB 070518 are shown in Fig. 1.
Galactic extinction (Schlegel et al. 1998)
\footnote{http://irsa.ipac.caltech.edu/applications/DUST/} is
corrected for all $V$,$R$ and $I$ bands data, which corresponds to a
reddening of $E_{B-V}=0.017$ mag in the direction of the burst. The $R$-
band data could be corrected for the flux contamination of its host
galaxy, whose magnitude is estimated to be $R=23.8\pm0.1$ mag
(Garnavich et al. 2007b).
The contamination is also interpreted as a combination of its host galaxy
and supernova bump (Dai et al. 2008).
As shown in Fig. 1, the $R$ band light
curve corrected for the flux contamination is labeled with "after",
while the $R$ band light curve without the correction is labeled
with "before". The
TNT data before 10$^{4}$ s are not affected
by the contamination very much.
Moreover,
the "after" $R$ band light curve of whole data (including TNT data and GCN data)
could be well fitted with a
simple power law with a decay index of $\alpha_{o}=0.87\pm0.02$
$(\chi^{2}/dof=54.6/17)$. The result is consistent with the fitting
index of $\sim$0.81$\pm$0.08 $(\chi^{2}/dof=14/10)$ only for TNT data.
The general trends of light
curves for the $R$ (labeled "before") and $I$ bands are
similar; however, the fluctuations exist in both light curves, which
might be a result of a low S/N ratio.

\begin{figure}
\centering
\includegraphics[angle=0,width=0.5\textwidth]{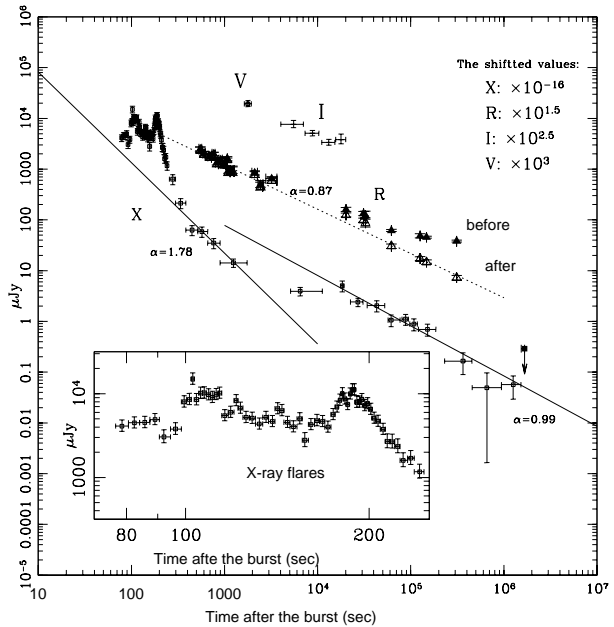}
\caption{Multiband light curves of GRB 070518.
Chromatic decay occurs between the X-ray and optical light
curves. The energy range of the X-ray flux is $0.3-10$
KeV. All $R$, $V$ and $I$ band data have been corrected for the Galactic
extinction. Solid and open triangles present the $R$ band data before and
after the correction of the flux contamination of the host galaxy,
respectively.
The insert  shows the flares in the X-ray
light curve at the beginning.
 }
   \label{Fig:lc}
   \end{figure}


In order to investigate the decay properties between optical and
X-ray light curves, we use the X-ray afterglow data downloaded from
the website of the UK {\em Swift} science data center
\footnote{http://www.swift.ac.uk/xrt\_curves/00279592/}(labeled as
USS) (Evans et al. 2007), where some analysis results are given,
including the spectral indices for each segments in the X-ray light
curve. These spectral indices are used directly for our later
analysis. The X-ray light curve is also plotted by squares in Fig. 1, 
showing several drastic flares from the
beginning up to about 250s after the burst trigger. 
Considering the
sparse data between 2000 and $3\times10^{4}$ sec, we fit the light
curve after 250 sec by separating it into two parts. one is between
250 sec and 2000 sec, and the other is from $3\times10^{4}$ sec to
over $6\times10^{6}$ sec. The first part is well fitted by a single
power law with an index of $\alpha_{x,1}=1.78\pm0.11$
($\chi^{2}/dof=0.76/3$), while the second is well fitted by a single
power law with a slope of  $\alpha_{x,1}=0.99\pm0.08$
($\chi^{2}/dof=5.83/8$). As shown in Fig. 1, the extrapolation of
the first fitting is below the data detected after 1000 sec,
suggesting that there is energy injection (Zhang et al. 2006)
between the two segments, shown as a shallow decay. Therefore, the
X-ray light curve displays a "steep-shallow-normal" decays
(Zhang et al. 2006).


\subsubsection{First Decay in X-ray Light Curve}

The first steep decay in the X-ray afterglow light curve
is usually explained by the
Curvature Effect (e.g. Kumar \& Panaitescu 2000; Zhang et al. 2009).
According to their model, temporal index $\alpha$ and spectral index
$\beta$ have a tight relation $\alpha=2+\beta$.
For GRB 070518, the first
slope $\alpha_{\mathrm{x,1}}$ (1.78) of the X-ray light curve
is definitely lower than the value (3.22, from  $2+\beta$)
expected for high-latitude emission (Here, $\beta=1.22$ from USS).
However, the high-latitude effect only provides
an upper limit to the decay slope,
as it assumes that the shell emission stops abruptly after
the initial pulse (Mangano et al. 2007).
The decline could be made in two cases:
one is that residual emission from the shocked shells
is still present, in which,
the high-latitude emission would contribute only a small
fraction of the overall radiation;
the other is that the
X-ray band is below the cooling frequency, in which case, the
expected decay slope of high-latitude radiation would naturally be
shallower than 2.
For GRB 070518, the first possibility might be more reasonable,
from the analysis for other segments of the X-ray light curve in the later sections.


\subsubsection{Other Segments}

The X-ray afterglow is analyzed by applying the closure relations given
by Zhang et al. (2006) to the X-ray spectral and temporal indices.
These relations help us to determine the location of the observed
X-ray band relative to the synchrotron frequencies
($\nu_a$,$\nu_m$,$\nu_c$) and the environment where the burst
occurred.

The normal decay segment with $\alpha_{x,3}=0.99$ and
$\beta_{x,3}=1.11\pm0.11$, agrees with the non-injected closure
relations: ${\beta={p\over2}}$ and ${\alpha={3p-2\over4}}$. These
relations are consistent with two cases
: when $\nu_{x}>\nu_m$ in fast cooling phase or
when $\nu_{x}>\nu_c$ in slow cooling phase. Also, The environment condition
[interstellar medium (ISM), Wind] could not be distinguished for the 
above cases.

The application of the closure
relations to the X-ray temporal and spectral index in normal decay segment,
combining the analysis in $\S$ 3.1 for the X-ray afterglow
during 2000$\sim$2$\times10^{4}$sec, suggests that
there is energy injection before the start of the normal decay phase.


\subsubsection{Flares in the X-ray Afterglow}

The X-ray light curve of GRB 070518 shows drastic flares between 80 and 200s
since the trigger, as shown in the insert in the Fig. 1.
Several GRBs were reported to have flares in X-ray afterglows (e.g.
Bernardini et al., 2009; Krimm et al., 2007b).
With the method of Krimm et al. (2007b), the
ratios of raise time duration and the middle time for two big flares
(peak times of about 106.9 s and 189 s, respectively) are calculated.
$\Delta$T/T for two flares was about 0.2 and 0.1, respectively, and
the $\Delta$F/F was about 1. Such steep and large amplitude rises
in the observed flux, as in GRB 070518, could not be explained by a
sudden increase in the external density (Nakar \& Granot 2007).
Another model, a patchy shell model, can produce the flare with
$\Delta$F/F of about 1 (Nakar \& Oren 2004), athough it cannot
interpret the flares with $\Delta$T/T $<<$ 0.1. For two big flares
of GRB 070518, $\Delta$T/T are about 0.2 and 0.1,respectively.
Therefore, the model might be possible. Moreover, such dramatic
flares are also related to refreshed shocks or the central engine activity
(e.g. Zhang et al. 2006).
This is more reasonable for the case of GRB 070518,
if we consider the
above analysis that energy injection was still occurred after these flares.


\subsubsection{ Chromatic Decay and Afterglow Model}

Chromatic decay could be seen in the case of GRB 070518, as shown in
Fig. 1.
The X-ray afterglow shows a "steep-shallow-normal" decay
(Zhang et al. 2005), while optical afterglow is decaying
with a constant index of 0.87 (if the correct of the
late contamination was right).
The decay index of the optical light curve
($\alpha_{opt}$=0.87) is slightly smaller than that of the normal decay segment
(0.99) of the X-ray afterglow.
The normal decay segment is usually explained the external
shock model (Sari et al. 1998), as expected for the optical afterglow.
However, the origin of the normal phase might also be related
to the internal energy dissipation (see as the discussion
by Urata et al. 2007).
Thus, the classical external shock model will tested with the burst.
According to the model, the difference between the decay indices
of the optical and X-ray afterglow should be between -0.25 for the
uniform ISM case and 0.25 for the wind medium case (Urata et al. 2007).
For GRB 070518, $\alpha_{\rm o}-\alpha_{\rm X}=0.87-0.99=-0.12$, ignoring the
uncertainties of those values.
The case of GRB 070518 agrees with the prediction of
the classical external shock model.
Moreover, with the relation derived by Urata et al. (2007) :
$\alpha_{\rm o} - \alpha_{\rm X} = -0.25 + s/(8-2s)$,
where s shows the condition of ambient matter density as a formate of $n<r^{-s}$,
the value of s for GRB 070518 could be deduced to be 0.84. This
suggests that the surrounding matter of the burst is a
mix of ISM  and wind-type medium.


\subsection{Testing the Amati and Ghirlanda relations}

The time-averaged spectrum of GRB 070518 from T-1.8s to T+4.5s is
best fitted by a simple power-law model with a photon index
$\Gamma=2.11\pm0.25$ (Krimm et al. 2007a; Guidorzi et al. 2007), corresponding
to the energy spectral index of 1.11$\pm$0.25 ($\beta=\Gamma-1$).
The fluence in the 15-150 KeV band is 1.6$\pm$0.2$\times$10$^{-7}$ erg cm $^{-2}$
(Krimm et al. 2007a), which is lower than the average value of the
{\em Swift} burst (Sakamoto et al. 2008a).

\subsubsection{Amati relation}

Fig. 2 shows the relation between $E_{peak,i}$ in the rest frame
and $E_{iso}$ (i.e. Amati relation; Amati et al. 2002). 
The data shown by open symbols and the best fitting
parameters are obtained from Amati et al. (2008).
For GRB 070518,
spikes in the BAT light curve reported by
Guidorzi et al. (2007) were only clearly seen under 50 KeV, which
implies that $E_{peak,o}$ might be less than 50 KeV. Besides,
if $E_{peak,o}$ is within the BAT energy range, the photon
index $\Gamma$ of a simple power-law fit and $E_{peak,o}$ are well
correlated with a relationship (Liang \& Zhang 2005):

$logE_{peak,o}=(2.76\pm0.07)-(3.61\pm0.26)log\Gamma$\\
Sakamoto et al.(2009) also obtained a similar conclusion by
combining the simulation:
$logE_{peak}=3.258-0.829\Gamma$\\
With $\Gamma=2.11$ and the two relations above,
$E_{peak,o}=38.8^{+16.96}_{-11.52}$ and 30.8 are obtained,
respectively.
The mean value of $\sim$ 35 KeV is taken from the above calculations
as its peak energy in the observational frame.
Thus $E_{peak,i}=(1+z) \times E_{peak,o}\sim60$ KeV in the rest frame.
As a result, GRB 070518 is consistent
with Amati relation within 3 $\sigma$ uncertainties, as shown in Fig. 2.

\begin{figure}
\centering
\includegraphics[angle=0,width=0.5\textwidth]{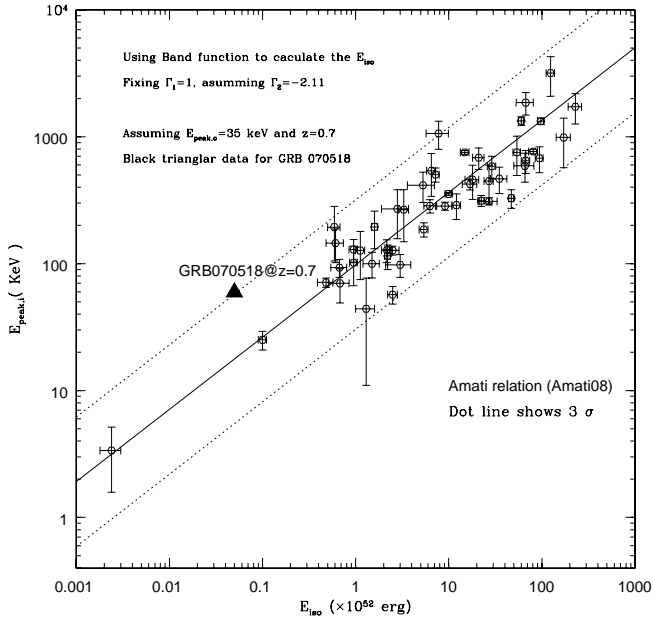}
\caption{ Correlation of E$_{peak,i}$ and E$_{iso}$ ( the Amati
relation). The solid triangle is shown for GRB 070518 when $z=0.7$ and
E$_{peak,o}=35$ KeV. E$_{iso}$ is calculated using the Band
function (Band et al. 1993), assuming photon index $\Gamma_{2}=2.11$ of the simple power law is
the higher photon index, and
fixing $\Gamma_{1}=1$ as the lower photon index.
The data shown as open symbols are taken from Amati et al. (2008). 
} \label{Fig:demo1}
\end{figure}

\subsubsection{Ghirlanda relation}

Ghirlanda, Ghisellini \& Lazzati (2004) proposed a relation between peak energy
$E_{peak,i}$ in the rest frame and the collimation-corrected energy
$E_{\gamma}$:
$E_{peak,i} \simeq 480\times (E_{\gamma,51})^{0.7}$
KeV,
where $E_{\gamma,51} = E_{\gamma}/10^{51}$ergs and
$E_{\gamma}=(1-cos\theta)E_{\gamma,iso}$.
In this relation, the jet angle $\theta$ is very important,
which was given by Sari, Piran \& Halpern (1999):\\
$\theta = 0.161\times (t_{jet,d}/(1+z))^{3/8} \times
(n\times\eta_{\gamma}/E_{\gamma,iso,52})^{1/8}$,\\
where z is the redshift, $t_{jet,d}$ is the break time in days. For
GRB 070518, the time of a possible jet break should be later than
the last observation of X-ray afterglow, $t_{jet,d}>10^{6}$ sec,
according to the analysis of $\S$ 3.1. Thus,the opening angle of jet
$\theta$ should be larger than 25 degree, under an assumption of the
typical values: $n=3$ cm$^{-3}$ and $\eta_{\gamma}=0.2$ (Ghirlanda
et al. 2004). From the values of $\theta$ and E$_{\gamma,iso,52}$,
$E_{\gamma}$ is estimated to be larger than 6$\times10^{49}$ ergs.

However, $E_{peak,i}$ has been
estimated to be about 60 KeV in the observation frame,
as analyzed in $\S$3.2.1. The $E_{peak,i}$ gives an
$E_{\gamma}$ of about 5.1$\times10^{49}$, based on the Ghirlanda relation.
This value is slightly less than that from the above analysis (6$\times10^{49}$ ergs),
but the values are consistent with each other, 
considering all the assumptions we have made.
Therefore,
GRB 070518 might agree with Ghirlanda relation from above analysis.


\subsection{One of the Optically Dimmest Long Bursts to Date}

Nardini et al.(2006) found that GRB optical luminosity at 12 hours
(at rest-frame) after the trigger shows a bimodal distribution. With
a redshift of 0.7 for GRB 070518, 12 h at the rest frame
corresponds to 20.4 hours in the observer frame.
Following the method
of Nardini et al. (2006), and assuming $\beta_{o}=0.8$,
$L_{opt,12h,i}=2\times10^{28}$ erg s$^{-1}$ is obtained
based on the redshift of 0.7, and $\alpha_{o}=0.87$.
The low value of $L_{opt,12h,i}$ of GRB 070518 means that the burst is 
one of the optically dimmest GRB afterglows so far (Nardini et al.
2008).

We have also noted that the result might be affected by the subtraction
of its contamination at late phase. In order to investigate this, 
we just calculate the optical luminosity at the rest frame
with the data without correction of the contamination of its
possible host galaxy. In this case, $L_{opt,12h,i} \sim
6\times10^{28}$ erg s$^{-1}$ is obtained with the assumptions as
above, indicating that the burst is really an optically dim burst.
Moreover, if the real redshift of the burst was lower than 0.7, its
luminosity would become much lower than the value given above.

Another method to investigate the optically dim luminosity of GRB
070518 is to compare its light curve with other long bursts
directly. Kann et al. (2007) have collected GRBs in the pre-{\em
Swift} and {\em Swift} era to compare the luminosity of GRBs in the
common redshift of 1. In this sample,  GRB 050416A, GRB 060512 and
GRB 070419A are at the lower edge of the distribution of optical
light curves (Seen in the Figure 1 of Kann et al. 2008). In order to
make the comparison, we have collected $R$ band data of the above three dim
bursts from GCN Circular and other literature.
These data are transformed into a common redshift $z=1$, with
$\beta_{opt}=1.3$ for GRB 050416A (Soderberg et al. 2007) and fixed
$\beta_{opt}=0.8$ for other bursts, GRB 070518, GRB 060512 and GRB
070419A. In the transformation, all magnitudes are corrected for the
Galactic extinction, and without considering the extinction from
their host galaxies. In addition, for GRB 070518, flux contribution
of host galaxy is corrected. As shown in Fig.3, the luminosity of
GRB 070518 is close to that of the three bursts, based on the redshift
of 0.7. Therefore, GRB 070518 should belong to optically dim
bursts, like the other three bursts (e.g. GRB 050416A, GRB
060512 and GRB 070419A).


\begin{figure}
\centering
\includegraphics[angle=0,width=0.5\textwidth]{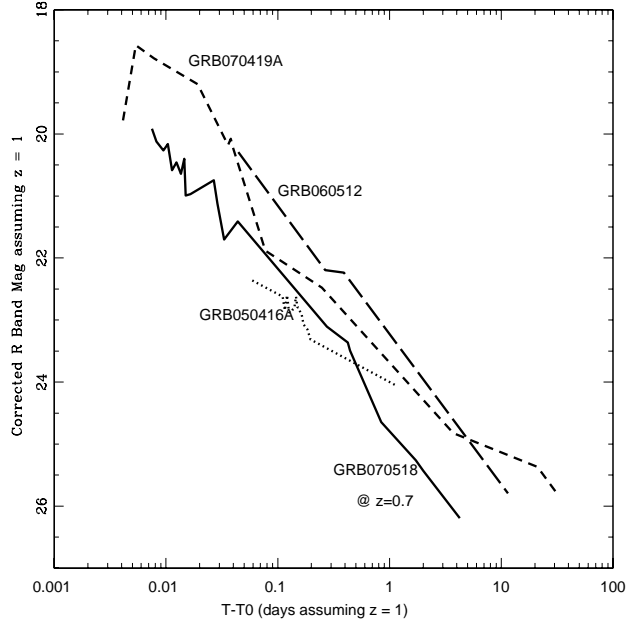}
\caption{Comparison of $R$ band luminosity of GRB 070518 with three other
long bursts after transformation into the same redshift of $z=1$.
The comparison long bursts are GRB 050416A, GRB 060512A and GRB
070419A, as labeled in the plot. GRB 070518 was shown as a solid
line based on the redshift of $z=0.7$.
The reference for data are as follows: GRB 050416A (Soderberg et al. 2007); GRB
060512A (Mundell et al. GCN5118; Malesani et al. GCN5122;
Cenko GCN5125; Milne GCN5127); GRB 070419A (Cenko et al. GCN6306; Williams et al.
GCN6328; Milne et al. GCN6341; Nishiura et al. GCN6308; Updike et al. GCN6317;
Lizuka et al. GCN6316; Fu et al.GCN6311;Pozanenko et al. GCN6407;
Garnavich et al.GCN6406; Hill et al. GCN6486).}
\end{figure}


\subsection{Extinction}

One of the interpretations for the optically dim luminosity is
the extinction and absorption from circumburst
medium in its host galaxy.
Here, we try to estimate
the extinction along the line-of-sight to the burst. The observed
neutral hydrogen column density along this line-of-sight to GRB
070518 determined by XRT onboard {\em Swift} is about
$N_{H}=(9.6\pm1.6)\times10^{20}$ cm$^{-2}$ (Guidorzi et al. 2007),
larger than the previously measured Galactic value
($N_{H}=2.2\times10^{20}$; Dickey \& Lockman, 1990). The excess
value is about $(7.4\pm1.6)\times10^{20}$ cm$^{-2}$. If the excess
value is from  a previously unobserved over-density located within
or close to our Galaxy, then the excess extinction $A_{V}=0.64$
would be obtained with the excess value of neutral hydrogen, based
on the relation found by Predehl \& Schmitt (1995):
$A_{V}=0.56\times N_{H}[10^{21} cm^{-2}]+0.23$ mag. If this is the
case, extinction should not be a prominent cause for the optically
dim luminosity.

It is likely that the excess column density is located within the
host galaxy of GRB 070518. As in the discussion for GRB 051022
(Nysewander et al. 2006), the rest frame $N_{H}$ is given by the
observed $N_{H,o}\times(1+z)^{2.6}$. The source frame is about
$N_{H,i}=(3\pm0.6)\times10^{21}$ with z=0.7. From the relation of
host extinction $A_{V,i}=A_{V,obs}\times(1+z)$ mag (Nysewander et
al. 2006), $A_{V,i}=3.2$ mag would be given in the rest frame based
on a redshift of 0.7.
Clearly, under these conditions, high extinction
would be a dominate cause of the optically dim luminosity of GRB
070518.

However, all above estimates are made with an assumption of the
Milky Way-type extinction. The optical extinction from the measured
$N_{H}$ absorption should depend on the assumed dust-to-gas ratio.
Some evidence shows that the extinction law in the GRB host galaxy
is likely to be a Small Magellanic Clouds (SMC) type. Using the small
dust-to-gas ratio of the SMC (Pei 1992), the host extinction
$A_{V,i}$ would be 0.3 mag when $z=0.7$. The difference of the
extinction value estimated above is very large, about a factor of
10. Actually, the real extinction law in the GRB host galaxy is
uncertain in most cases, like GRB 050401, for example, the
extinction $A_{V}$ is estimated to be $0.62\pm0.06$ by fitting the
optical spectrum, which is lower than the value $A_{V}\sim9.1$
inferred from the X-ray column density (Watson et al. 2006). One
reason for the case of GRB 050401 was that high energy release of
gamma-ray and hard X-ray photons might destroy the circumburst dust
in the GRB host galaxy (Watson et al. 2006). The reason might also
be the case of GRB 070518.


\subsection{Comparison with Other Three Optically Dim Bursts }

As shown in Fig.3, three {\em Swift} optically dim  bursts GRB
050416, GRB 060512A and GRB 070419A are labeled to compare with GRB
070518. In order to investigate the reason for the dim luminosity of GRB
070518, we try to compare them in different aspects.

Table 2 presented the main parameters for the three bursts
along with those of GRB 070518.
There are several similar properties, as follows.\\
(1) For all the bursts, T$_{90}$ is lower than 10s, except
for GRB 070419A. This means that they belong to the
shortest part of the "long GRB" (Paciesas et al. 1999),
or intermediate long GRB ( Horvath et al. 2008 ) classification.\\
(2) The photon indices of these bursts are larger than 2, which
implies that the spectra are so soft that these bursts
belong to XRRs and XRFs. \\
(3) The redshifts of all these bursts are lower than 1.

\begin{table}
\caption[]{A comparision of properties for four GRBs. $\Gamma$ is the photon index at 15-150 KeV. 
$F$ is fluencee $\times$10$^{-7}$ erg at 15-150 keV. $Z$ is redshift.\\}
  \label{Tab:publ-works}
  \begin{center}\begin{tabular}{lcccccl}
  \hline\noalign{\smallskip}
GRB &  $T_{90}$ &  $\Gamma$  &   F &  Z &  $A_{V}$ & XRR \\
name &  (s)   &            &     &    &        &  or XRF \\
  \hline\noalign{\smallskip}
070518   &  5.5  &  2.11  &  6.02  &  0.7$\sim$1   &  --  &    yes  \\
050416A$^a$  &  2.4  &  3.4   &  3.5   &  0.6535       &  0.2(0.32) &  yes  \\
060512$^b$   &  8.6  &  2.49  &  2.3   &  0.4428(2.1)  &  --   & yes   \\
070419A$^c$  &  116  &  2.35  &  5.6   &  0.97         &  --   &  yes   \\
 \noalign{\smallskip}\hline
  \end{tabular}\end{center}
\begin{list}{}{}
\item[]
$^a$ The references for GRB 050416A are Sakamoto et al.(2005,2006), Holland et al. (2007) and  Kann et al. (2007).\\
$^b$ The references for GRB 060512 are Cummings et al. (2006), Starling R. et al. (2006),
Bloom et al. (2006) and Kann et al.(2007).\\
$^c$ The references for GRB 070419A are Cenko et al. (2007) and Stamatikos et al. (2007).\\
\end{list}
\end{table}

Another aspect concerns the absorption and extinction of the GRB host
galaxy. Of the bursts mentioned above, only  GRB 050416A  has
been studied extensively (e.g. Sakamoto et al. 2006; Mangano et al.
2007; Soderberg et al. 2007), and the extinction from  its host
galaxy has been estimated to be $A_{\rm V}=0.2$ (Holland et al. 2007),
or $A_{v}=0.32\pm0.17$ (Kann et al. 2007). The two values are
similar to the average value ($A_{\rm V}$=0.2) of GRBs based on the
study of Kann et al. (2007, 2008). However,  they are significantly smaller than the
extinction (A$_{\rm V}\sim3.8$) expected from the hydrogen column
density inferred from X-ray observations of XRF 050416A, assuming a
dust-to-gas ratio similar to that found for the Milky Way
(Holland et al. 2007). The low value of extinction from  host galaxy
of GRB 050416A implies that the extinction of host galaxy of GRB
070518 might be not large, only about $A_{\rm V}=0.3$, as analyzed
in $\S$3.4.


\subsection{Optical to X-ray Energy Spectral Index}

Optically dim bursts might be related to dark bursts (Nardini et
al. 2008). Considering that GRB 070518 is an optically dim burst, we
wonder whether the burst is a dark burst or not.

The index of optical to X-ray spectrum $\beta_{\mathrm{ox}}$ ($<0.5$) at 11
h after the burst trigger is used to define dark bursts
(Jakobsson et al. 2004).
Cenko et al. (2009) argued that the X-ray flux at about 11
hours might be contaminated by the shallow decay for some bursts. As
a result ,the X-ray flux would be larger than the real X-ray flux
radiated along with optical afterglow, which would make some normal
bursts  "dark" bursts. They calculated $\beta_{\mathrm{ox}}$ for
the bursts in their catalogue at 1000s after the trigger time.
However, it is too early for most of the {\em Swift} bursts to carry out the
calculation at 1000s after the trigger, because the X-ray light curves
in some {\em Swift} GRBs might be at the steep  or
shallow decay segment about this time.

For GRB 070518,  $\beta_{\mathrm{ox}}$ at 1000s, 11
hours and $10^5$s after the burst are calculated by interpolation,
as shown in Fig.4.  $\beta_{\mathrm{ox}}$ for GRB 070518 is
0.666, 0.606 and 0.543 at the above three epochs, respectively.
It seems that the burst is not a dark burst, for
all of
$\beta_{\mathrm{ox}}$ are slightly larger than 0.5.
However, spectral index
$\beta_{ox}$ depends on the energy distribution $p$ of electrons
(Jakobsson et al. 2004), For GRB 070518, $p$ should be 2.22$\pm0.22$, as $\beta_{x}=1.11\pm0.11$
(from analysis of USS) and $p=2\beta$. In this case, the criterion for dark bursts should
become $\beta_{\mathrm{ox}}<0.6\pm0.1$. Consequently,
GRB 070518 might be a
dark burst (Jakobsson et al. 2004) or a gray burst (Zheng et al. 2009).

\begin{figure}
\centering
\includegraphics[angle=0,width=0.5\textwidth]{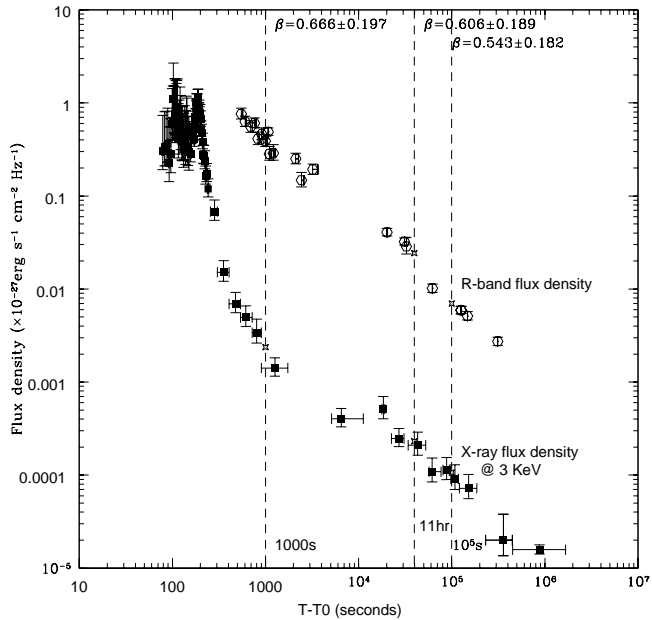}
\caption{ The optical and X-ray flux density light curves. The optical 
corresponds to the $R$ band, shown as
open hexagonal. The X-ray data at 3 KeV is shown as
solid squares. The spectral indices are calculated at different times
(1000s, 11 hours and $10^5$s) by the interpolation of their light
curves. The optical to X-ray spectral indices $\beta_{ox}$ are also
shown.
 }
\label{Fig:demo4}
\end{figure}


\section{Summary and Conclusion}

We have presented optical afterglow observations of GRB 070518 
from the TNT and
an analysis of the optical and X-ray light curves.
The optical afterglow shows a constant power law decay $\sim$0.87,
while the X-ray shows a "steep-shallow-normal" decay behavior.
Based on the upper limit of redshift of 0.7, the optical
luminosity at 12 hours in the rest
frame was estimated to be $2\times10^{28}$ erg s$^{-1}$,
which means that the burst is an optically dim burst.
The conclusion is also supported by comparing with three other optically dim
bursts: GRB 050416A, GRB 060512 and GRB 070419A. The spectral
indices of $\beta_{\mathrm{ox}}$ at 1000 s, 12 hours,10$^{5}s$ are
calculated. All $\beta_{\mathrm{ox}}$ were slightly larger
than 0.5, indicating that GRB 070518 might be a dark burst or a gray burst,
according to the definition of Jakobsson et al. (2004). The optical
extinction in its host galaxy inferred from  X-ray hydrogen column
density is 3.2 with Galactic extinction law, and
0.3 with SMC extinction law.

Afterglow model has been applied to the light curves. Energy
injection is found to occurr for the X-ray afterglow before the
start of the normal decay segment. The behavior of the normal
segment of the X-ray afterglow and the optical afterglow is consistent
with the prediction of the classical external shock model (Urata et
al. 2007), indicating that they come from the same origin. As an
XRR burst, GRB 070518 agrees with Amati relation. Moreover, it
fills the gape of high-luminosity and low-luminosity bursts in the Amati
relation, similar to GRB 050416A (Sakamoto et al. 2005). It is also found that 
GRB 070518 agrees with the Ghalanda
relation within
reasonable assumptions.

Moreover, a comparison with three optically dim bursts
(GRB 050416A, GRB 060512 and GRB 070419A and GRB 070518)
reveals that several similar properties were shared:
burst duration T$_{90}$$<$10s (expect for GRB 070419A, $T_{90}=116s$), soft
spectrum at BAT band $\Gamma$$>2$, low redshift $z<$1, etc.

\section*{acknowledgements}
We are very grateful to the anonymous referee, who provided excellent advice and comments,
and many suggestions for improving the paper.
We wish to thank Dr.Xinyu Dai for the discussion about the
dark burst and the redshift of the burst.
Thanks go
to Dr.Yufeng Mao, Xuhui Han, Huali Li for their
kindly help in correcting and improving the writing. 
This work made use of
data supplied by the UK {\it Swift} Science Data Center at the
University of Leicester. This study has been supported by National
Basic research Program of China-973 Program 2009CB824800 and
by the National Natural Science Foundation of China-Grant
No.10673014 and 10803008.

\bibliographystyle{mn}

\label{lastpage}
\end{document}